\documentclass[preprint,preprintnumbers,amsmath,amssymb]{revtex4}

\usepackage{graphicx}
\usepackage{dcolumn}
\usepackage{bm}
\usepackage{multirow}
\usepackage[dvipsnames]{xcolor}
\usepackage{color}
\usepackage[normalem]{ulem}
\usepackage{amsmath}
\usepackage{gensymb}

\usepackage[draft]{changes}

\usepackage[colorlinks=true]{hyperref}

\begin{document}

\title{Probing the optical near-field interaction of Mie nanoresonators\\ with atomically thin semiconductors}

\author{Ana Estrada-Real$^{1,2}$}
\author{Ioannis Paradisanos$^{3,2}$}
\author{Peter R. Wiecha$^{4}$}
\author{Jean-Marie Poumirol$^{5}$}
\author{Aurelien Cuche$^5$}
\author{Gonzague Agez$^{5}$}
\author{Delphine Lagarde$^{2}$}
\author{Xavier Marie$^{2}$}
\author{Vincent Larrey$^{6}$}
\author{Jonas M\"uller$^{4}$}
\author{Guilhem Larrieu$^4$}
\author{Vincent Paillard$^5$}
\author{Bernhard Urbaszek$^{1,2}$}

\affiliation{\small$^1$Institute of Condensed Matter Physics, Technische Universität Darmstadt, 64289 Darmstadt, Germany}
\affiliation{\small$^2$Universit\'e de Toulouse, INSA-CNRS-UPS, LPCNO, 135 Avenue Rangueil, 31077 Toulouse, France}
\affiliation{\small$^3$Institute of Electronic Structure and Laser (IESL), Foundation for Research and Technology-Hellas (FORTH), 70013, Heraklion-Crete, Greece}
\affiliation{\small$^4$LAAS-CNRS, Universit\'e de Toulouse, 31000 Toulouse, France}
\affiliation{\small$^5$CEMES-CNRS, Universit\'e de Toulouse, Toulouse, France}
\affiliation{\small$^6$CEA-LETI, Universit\'e Grenoble-Alpes, Grenoble, France}

\begin{abstract}
Optical Mie resonators based on silicon nanostructures allow tuning of light-matter-interaction with advanced design concepts based on CMOS compatible nanofabrication. 
Optically active materials such as transition-metal dichalcogenide (TMD) monolayers can be placed in the near-field region of such Mie resonators. 
Here, we experimentally demonstrate and verify by numerical simulations coupling between a MoSe$_2$ monolayer and the near-field of dielectric nanoresonators. 
Through a comparison of dark-field (DF) scattering spectroscopy and photoluminescence excitation experiments (PLE), we show that the MoSe$_2$ absorption can be enhanced via the near-field of a nanoresonator. We demonstrate spectral tuning of the absorption via the geometry of individual Mie resonators. We show that we indeed access the optical near-field of the nanoresonators, by measuring a spectral shift between the typical near-field resonances in PLE compared to the far-field resonances in DF scattering. 
Our results prove that using MoSe$_2$ as an active probe allows accessing the optical near-field above photonic nanostructures, without the requirement of highly complex near-field microscopy equipment.
\end{abstract}

\maketitle


\textbf{Introduction}.--- Optical resonators are essential in many applications such as laser systems and sensing. The physical size and properties of the resonator are adapted to the specific application and to the relevant part of the electromagnetic spectrum \cite{novotny2012principles}. Optical resonators that are capable of amplifying optical fields in very small nanoscopic volumes, for addressing individual nanocrystals or molecules in the near-field, are called nanoresonators  \cite{kleemann2017strong, petric2022tuning,liu2016strong,zhou2017probing,sortino2019enhanced, bidault2019dielectric, mupparapu2020integration, brongersma2021road,chen2017enhanced, bucher2019tailoring, cihan2018silicon,shinomiya2022enhanced}. They can be fabricated by bottom-up techniques, such as growth of metallic nanoparticles, or top-down approaches, such as Si-nanoresonators on CMOS compatible substrates \cite{zhao2009mie, wiechaEvolutionaryMultiobjectiveOptimization2017,won2019into, gonzalez2021scaling, kuznetsov2016optically,kallelPhotoluminescenceEnhancementSilicon2013}.
Whereas the resonance energies of a resonator with macroscopic dimensions, such as a laser cavity, are directly accessible in a standard optical far-field measurement, the situation for nanoresonators is more challenging. 
It has been shown experimentally and in a substantial body of theory work that there is a shift between the optical resonance energy in the near-field compared to the measured resonance energies in the far-field. The exact near-field resonance is key for applications for example in sensing of molecules directly placed in the near-field \cite{chen2019modern} and it has been accessed up to now either in sophisticated tip-enhanced experiments or through extrapolation from far-field data \cite{messingerLocalFieldsSurface1981, rossComparisonFarfieldMeasures2009, katsEffectRadiationDamping2011, zuloagaEnergyShiftNearField2011, alonso-gonzalezExperimentalVerificationSpectral2013}.\\
\indent Here we show that by placing an atomically thin semiconductor directly in the near-field of individual dielectric nanoresonators, we have access to the near-field resonance energies without the use of complex near-field spectroscopy techniques. We compare the near-field resonance energies with far-field resonance energy measurements on the same resonators and observe a clear blue-shift of the near-field energies in our far-field results. We show tuning of the optical absorption of the atomically thin semiconductor MoSe$_2$ through the interaction with the nanoresonator near-field and our results are well reproduced by model calculations of the shift between near-and far-field resonances.

\begin{figure*}
\includegraphics[width=0.9\linewidth]{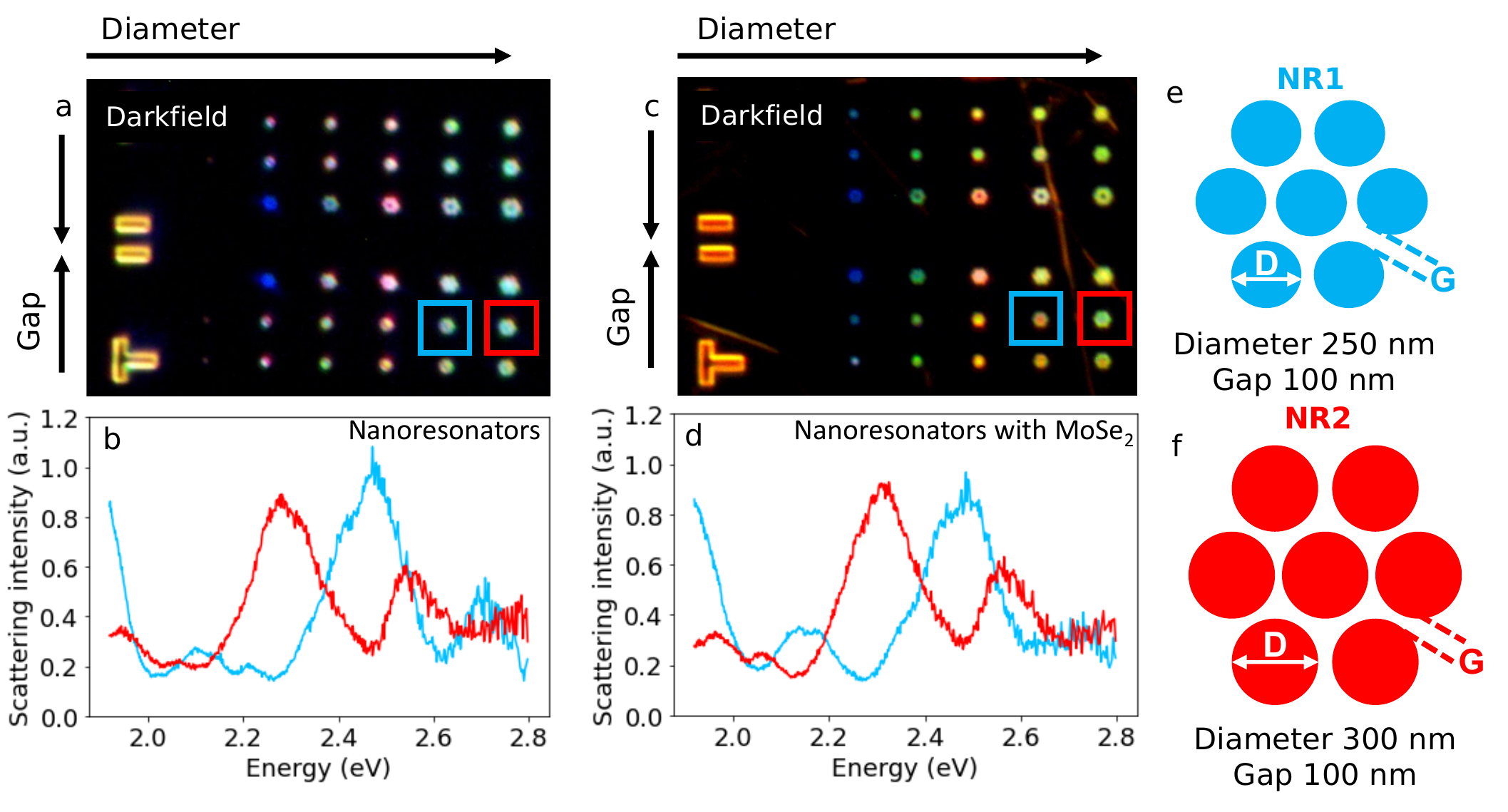}
\caption{\label{fig:fig1} \textbf{dark-field scattering images and spectra with and without MoSe$_2$} (a) dark-field microscope image of SiO$_2$/Si nanoresonators, the 15 structures on the top part of the image are hexamers (six pilars) with diameters varying from D=100-300 nm from left to right and the gap between pillars varying G=50-100-300 nm from top to bottom. The 15 structures on the bottom are heptamers (seven pillars) with the same variation. (b) Dark-field Scattering intensity of the two nanoresonators highlighted. (c) dark-field microscope image with a MoSe$_2$ monolayer on top. Caution, automatic white-balance was used, colors do not directly compare to~(a). (d)  Dark-field Scattering intensity of the two nanoresonators after transfer of MoSe$_2$ monolayer. (e and f) Sketches of the two selected nanoresonators \textcolor{cyan}{NR1} and \textcolor{red}{NR2}, respectively (top view - see supplement for side view).}
\end{figure*}
\textbf{Results and Discussion}.--- 
We fabricated two sets of Si/SiO$_2$ nanoresonator arrays on silicon-on-insulator (SOI) substrates. 
The nanoresonators are cylindrical pillars, arranged as close-packed heptamers as sketched in Fig.~\ref{fig:fig1}e and~\ref{fig:fig1}f. 
The diameters of the individual cylinders increase from 50~nm to 300~nm with steps of 50~nm. 
The gaps between neighbouring pillars are 50~nm, 100~nm or 300~nm.
The structures are fabricated as arrays of seven discs to match approximately the size of a diffraction limited, focused laser beam, in order to maximize the experimental signal associated with the Si-NRs.
On the final nanostructures, MoSe$_2$ monolayers are aligned and transferred on top of the nanoresonators using a micromanipulator system \cite{castellanos2014deterministic}.
A description of the fabrication process is given in the supporting information. 

In figure~\ref{fig:fig1}a we show dark-field (DF) images of the nanoresonators before MoSe$_2$ transfer, figure~\ref{fig:fig1}c shows the same sample after transfer of an MoSe$_2$ monolayer flake.
A bright-field microscope image of the nanoresonators after MoSe$_2$ deposition is shown in the supplement Fig. S3, where the MoSe$_2$ covered region is clearly visible.
By varying the Si-NR gap and diameter, different colors appear in the DF images, which demonstrates the geometry-dependent Mie resonance energy shifts of the individual resonators in the visible spectral range.

DF spectra are collected using the same setup as the DF images. To this end, the signal is sent to a spectrometer instead of the imaging camera. The setup for taking dark-field images and spectra is depicted in the supplement Fig.~\ref{fig:figS1}.
We will focus in the following on two selected nanoresonators, ``NR1'' (blue, Fig.~\ref{fig:fig1}e) and ``NR2'' (red, Fig.~\ref{fig:fig1}f). These heptamers have respective diameters of 250~nm (NR1) and 300~nm (NR2), the gap between the pillars is identical for both Si-NRs (100~nm).
We selected these two NRs because their main Mie resonances lie at different, well distinguishable energies.
We measured the scattered light intensity before and after the monolayer transfer, shown in Fig.~\ref{fig:fig1}b, respectively~\ref{fig:fig1}d. 
We observed a small, global blue-shift on the order of 10 meV for the resonances when the MoSe$_2$ monolayer was on top of the nanoresonators, this small energy-shift is at the limit of our detection accuracy. Knowing the size of this shift is helpful for analyzing the spectral shift between measured near-field and far-field resonances discussed below.\\

\begin{figure}
\includegraphics[width=0.99\linewidth]{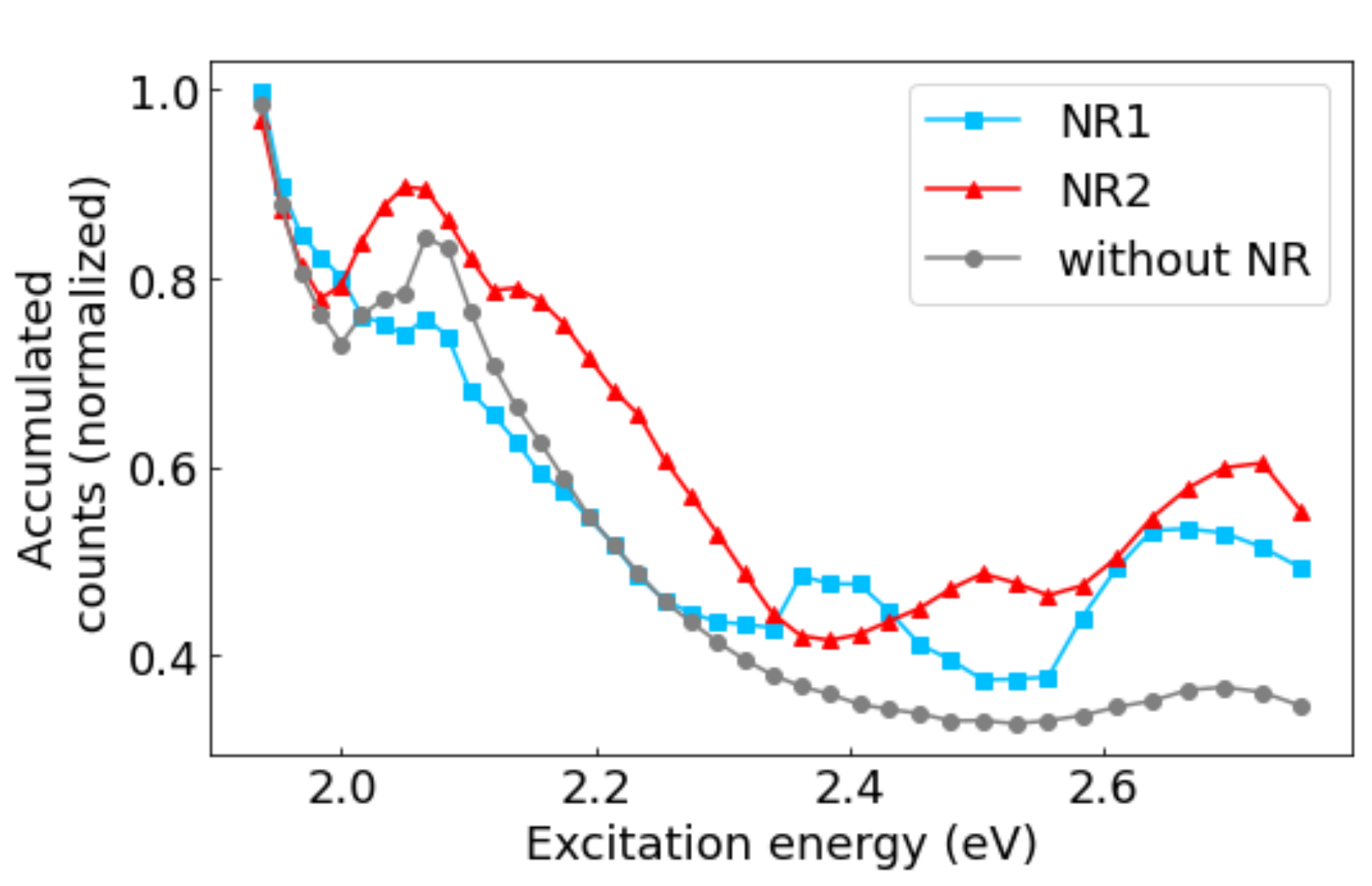}
\caption{\label{fig:fig2} \textbf{Photoluminescence excitation measurements.} Accumulated counts of the photoluminescence (PL) spectra measured on bare MoSe$_2$ (gray circles) and MoSe$_2$ on top of each nanoresonator NR1 (blue squares) and NR2 (red triangles), excited with a continuum laser at 40 different energies and at T=5~K.}
\end{figure}

\textit{Spectral shift between near-field and far-field.} 
In a second step, we want to analyze the impact of the nanoresonators on the absorption of the MoSe$_2$ monolayer. 
We carry out photoluminescence excitation (PLE) experiments \cite{shree2021guide} where temperature, position, and optical power are kept constant and the excitation wavelength is varied from 450~nm to 650~nm (corresponding to 2.75~eV - 1.90~eV). 
The optical power of the laser is set to 100~nW, after making sure that the MoSe$_2$ absorption is not saturated at this illumination power (see also supplemental figure~S4).
We avoid tuning the laser close to the MoSe$_2$ exciton emission peaks which occur around 745~nm ($\approx 1.66\,$eV)\cite{vanderzandeGrainsGrainBoundaries2013, cadizUltralowPowerThreshold2016, wang2015exciton}, in order to remain in a non-resonant excitation regime. During the entire measurements, the sample is kept in a closed-loop cryostat at a temperature of 5~K. 
Typical PL spectra are shown in the supplemental Fig.~\ref{fig:figS6}, where we also provide a comparison with room temperature measurements ($T=300$~K).
An illustration of the optical setup can also be found in the SI Fig.~\ref{fig:figS2}. 

\begin{figure*}
\includegraphics[width=0.96\linewidth]{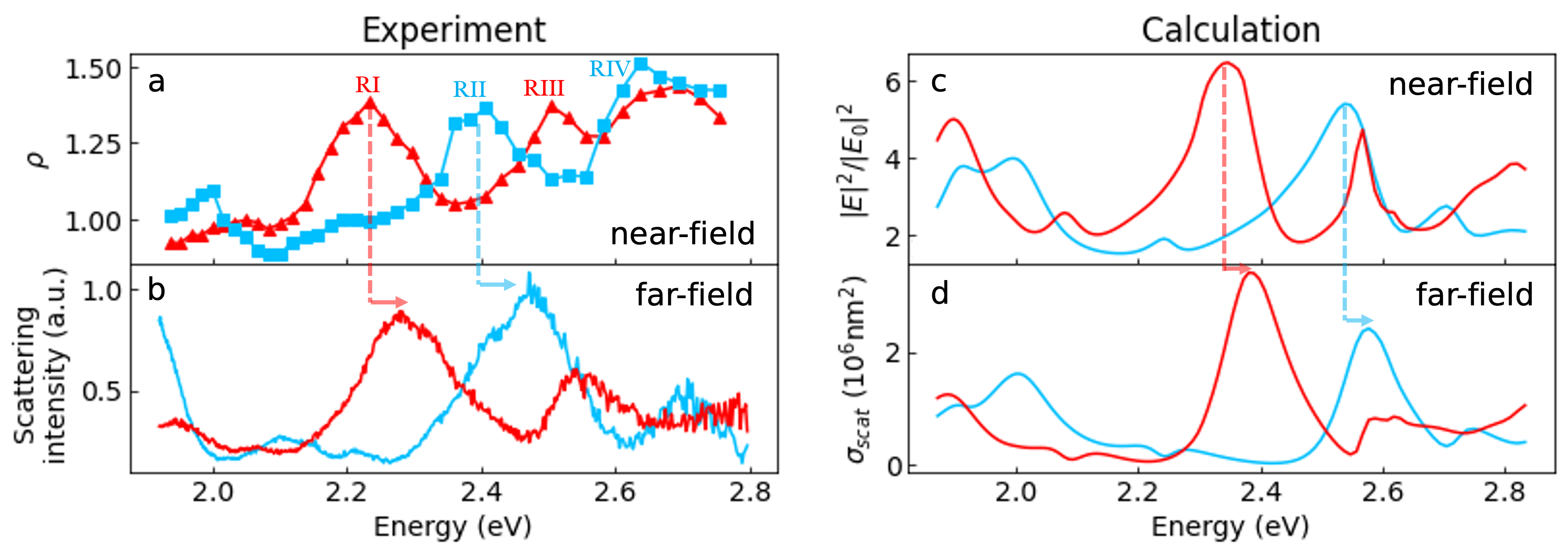}
\caption{\label{fig:fig3} \textbf{Tuning of resoncances and spectral shift.} Results of measurements and calculations for \textcolor{cyan}{NR1} and \textcolor{red}{NR2} (a) The result of dividing PLE over NR1 (blue squares) and NR2 (red triangles) by PLE on bare MoSe$_2$, where we plot the ratio $\rho$ as defined in Eq.~\ref{eq1}. (b) Dark-field scattering intensity, same data as Fig.~\ref{fig:fig1}b and (c) Calculated near-field intensity enhancement above the SiNRs, averaged in the plane of the TMD (d) calculated far-field intensity enhancement above the SiNRs, averaged in the plane of the TMD.}
\end{figure*}

The absorption and hence the PLE response of the bare MoSe$_2$ \cite{shree2021guide} is expected to vary with the excitation energy. This variation is a result of the material's band-structure, the presence of high energy excitonic states as well as coupling to phonons \cite{chow2017phonon, wang2015exciton, kozawa2016evidence, shree2018observation}. 
Our goal is to investigate how the absorption is modified by Mie resonances of the Si-NRs. 
To distinguish effects induced by Mie-resonances from material-related variations in the bare MoSe$_2$, we compare PLE measurements from  MoSe$_2$ monolayers lying on the flat substrate with MoSe$_2$ on top of the nanoresonators. 
This is shown in Fig.~\ref{fig:fig2} for NR1 and NR2. 
The PL emission spectra for each excitation laser wavelength are numerically integrated, by summation of the counts per second in the spectral range around the emission peaks, from 1.5\,eV - 1.7\,eV (see SI for raw spectra) \cite{atkinson1989introduction}. Hence, each data point in Fig.~\ref{fig:fig2} is a separate PL measurement for a specific laser excitation energy.
The direct comparison of the PLE measurements from MoSe$_2$ with and without Si-NR (c.f. Fig.~\ref{fig:fig2}), already reveals clear differences.
To better visualize the effect of the nanoresonators on the PLE signal, we divide the integrated PL intensities from MoSe$_2$ on the Si-NRs $I_{PL, \text{NR}}$ (blue squares and red triangles in Fig.~\ref{fig:fig2}) by the signal from the bare MoSe$_2$ $I_{PL, \text{MoSe$_2$}}$ (on flat substrate, gray dotted line in Fig.~\ref{fig:fig2}):
\begin{equation} \label{eq1}
\rho = \frac{I_{PL, \text{NR}}}{I_{PL, \text{MoSe$_2$}}} \, .
\end{equation}
This ratio $\rho$ gives an estimation of the absorption enhancement due to the presence of a nanoresonator supporting Mie resonances. The Mie resonances locally amplify the optical near-field, enhancing absorption in the monolayer which leads after carrier relaxation to a stronger PL emission.
Spectra of $\rho$ as a function of the excitation energy are shown in Fig.~\ref{fig:fig3}a. In order to show spectral tuning of the absorption enhanced by Mie-resonances we perform experiments on the two nanoresonators NR1 and NR2 (see Fig.~\ref{fig:fig1}e and f). We find that their main resonances (labelled RI, ..., RIV) occur at well separated energies (see table~\ref{Tab1} for values). 
Around the main Mie resonances we observe a PL emission enhancement of about 1.44. 
A comparison with the DF scattering spectrum (Fig.~\ref{fig:fig3}b) shows, that the PLE maxima are governed by the Mie-resonances. The spectral shifts between NR1 and NR2 are consistent between near- and far-field measurements. 
This is a clear signature that the MoSe$_2$ monolayer is coupled to the optical near-field above the Si-NR in the regime of weak coupling.

\begin{table}
\begin{center}
\caption{\label{Tab1} Energy of measured resonances, where we compare near-field (PLE) with far-field (dark-field scattering) experiments. Resonance labels according to Fig.~\ref{fig:fig3}a.}
\begin{tabular}{ |c|c|c|c|c| } 
\hline
 & RI (eV) & RII (eV) & RIII (eV) & RIV (eV) \\
\hline
PLE & 2.23 $\pm 0.05$ & 2.41 $\pm 0.05$ & 2.50 $\pm 0.05$ & 2.64 $\pm 0.05$ \\ 
dark-field & 2.28 $\pm 0.01$ & 2.46 $\pm 0.01$ & 2.54 $\pm 0.01$ & 2.71 $\pm 0.01$ \\ 
\hline
\end{tabular}
\end{center}
\end{table}

\textit{Theoretical simulations} : We compare the experimental PLE spectra to full-field simulations via the Green's dyadic method (GDM) using our own python implementation ``pyGDM'' \cite{wiecha2018pygdm, wiechaPyGDMNewFunctionalities2022}. The GDM is a volume discretization approach to solve Maxwell's equations in the frequency domain \cite{girardFieldsNanostructures2005}. 
A nanostructure of arbitrary shape and material is discretized on a regular hexagonal compact grid. We use tabulated refractive indices from literature for both parts of the pillars, the first 95~nm consisting of silicon \cite{edwardsSiliconSi1997} while the 30~nm thick capping is made of SiO$_2$ \cite{malitsonInterspecimenComparisonRefractive1965}.
With according Green's tensors \cite{paulusAccurateEfficientComputation2000}, we describe the layered substrate, where a bulk silicon substrate is followed by a SiO$_2$ spacer layer of 145~nm, on top of which the nano-structures are placed in air. 
We illuminate the system with a plane wave at normal incidence in the same wavelength range as used in the experiments and we incoherently average two orthogonal linear polarizations. 
We calculate the electric field intensity enhancement just above the SiO$_2$ capping on the Si pillars in an area of $1 \times 1$~\textmu m$^2$, corresponding approximately to the size of the focused Gaussian laser beam ($\approx$ emission beam diameter). Sketches and more details about simulations and model geometry can be found in the SI.

The simulated spectra of the average near-field enhancement above the top surface are shown in Fig.~\ref{fig:fig2}c for heptamers with pillar diameters of $D=250$~nm (NR1, blue) and $D=300$~nm (NR2, red), and the far-field scattered intensity for the same structures shown in Fig.~\ref{fig:fig2}d.
We observe a good agreement with the experimental data, the simulations reproduce in particular all major resonance features. Very importantly our simulations reproduce the red-shift of the resonances when the structure size increases from $D=250$~nm to $D=300$~nm and the near-to far-field shift.
Please note that the systematic blue-shift of around $0.05-0.1$\,eV between the absolute values of the simulated resonances and the experiment can be explained by the rough discretization of the Si discs' circular cross-sections on a coarse, regular mesh. 
Also, inaccuracies of the fabricated sample dimensions may play a role, in fact, deviations in the order of $20$-$25$ nm between design and actual sample would induce an according systematic shift \cite{patoux2021challenges}.

\textit{Discussion} : The comparison of the far-field DF scattering spectra with the MoSe$_2$ PLE spectra clearly demonstrate, that the Mie resonances of high-index dielectric nanoresonators can be used to tailor the absorption of a TMD layer. 
Furthermore, our results show that the PLE measurements access the optical near-field and probe the local field intensity enhancement. In the lateral directions (parallel to the substrate plane) the measurements reflect the average near-field intensity in a large zone, corresponding approximately to the illuminated area. Perpendicular to the surface on the other hand, the local near-field is probed in an extremely narrow vertical region, because the TMD is atomically thin. 

An unambiguous proof that we do indeed probe the optical near-field with the PLE technique is provided by the observation of a systematic shift between near-field spectra (PLE) and far-field measurements (DF scattering). 
This shift is of the order of $50$\,meV (see also Table~\ref{Tab1}) and is reproduced by our simulations (Fig.~\ref{fig:fig3}c and d).
It corresponds to the widely studied resonance shift between far-field and near-field, where the near-field amplitude is shifted to lower energies, compared to the far-field response. To get an intuitive, qualitative understanding, this shift can be explained with a damped harmonic oscillator (HO) model. The HO amplitude is given by:
\begin{equation}\label{eq:damped_oscillator}
    A(\omega) = \frac{A_{\text{drive}}}{\sqrt{\big(1-(\omega/\omega_0)^2\big)^2 + \gamma^2 \, (\omega / \omega_0)^2}} \, ,
\end{equation}
where $A_{\text{drive}}$ is the driving amplitude, in our case corresponding the amplitude of the incident field.
Solving Eq.~\eqref{eq:damped_oscillator} for its extrema, one finds easily that the damping term $\gamma$ leads to a red-shift of the maximum resonator amplitude $\omega_{A_{\text{max}}}$ with respect to the eigenfrequency $\omega_0$ \cite{zuloagaEnergyShiftNearField2011}:
\begin{equation}
    \omega_{A_{\text{max}}} = \omega_0 \sqrt{1 - \gamma^2/2} \, .
\end{equation}
On the other hand, far-field observables like the scattering or extinction cross section are proportional to the oscillator's kinetic energy, whose time average can be shown to be maximum at the non-shifted eigenfrequency~$\omega_0$ \cite{zuloagaEnergyShiftNearField2011}.
While the effect has first been discussed for lossy plasmonic nanoresonantors \cite{messingerLocalFieldsSurface1981}, it has been explicitly shown later that radiative damping, responsible for broadening of leaky Mie resonances like in our SiNRs, leads to an analogous near-field shift \cite{zuloagaEnergyShiftNearField2011}.
The shift has been experimentally demonstrated on plasmonic structures via scanning near-field microscopy (SNOM) measurements \cite{alonso-gonzalezExperimentalVerificationSpectral2013}.

Without complex SNOM equipment, here we experimentally observe a red-shift of the PLE spectra compared to the far-field scattering results. It is in the order of the shift predicted by our near-field enhancement simulations, and therefore a clear signature that we indeed probe the optical near-field of the silicon nanostructures.

These observations rely on a sufficient spectral resolution to observe the near-field red-shift in the PLE data. Please note that our approach also works at room temperature, as shown in the SI Fig.~\ref{fig:figS5}.

\textit{In conclusion,} we transferred monolayers of MoSe$_2$ on top of Mie resonant silicon nanostructure arrays and compared the far-field scattering of these scatterers, using DF spectroscopy, with spectrally resolved measurements of photoluminescence enhancement.
The latter is probing the optical near-field at the location of the TMD monolayer, i.e. just above the silicon nanostructures. 
We found that the PLE spectra show the same resonant features as the far-field measurements. 
We also observe the typical near-to-far-field spectral shift between the two types of measurements, which unambiguously confirms that we have access to the optical near-field of the silicon nanostructures. 
All results are confirmed by numerical full-field simulations.
Our results demonstrate that the absorption efficiency and consequently the emission of direct-bandgap monolayer semiconductors can be enhanced and spectrally tuned by placing optically resonant nano-structures in their vicinity. 
It furthermore proves that the PLE measurements provide access to the optical near-field in an extremely narrow vertical region in the order of a nanometer (thickness of the TMD monolayer).
Our work offers important insights for the design of efficient TMD-based nano-devices for light detection and emission.


\textbf{Methods}.--- The silicon pillars were fabricated in a top-down approach via electron-beam lithography (EBL) and subsequent anisotropic plasma etching \cite{han_realization_2011, guerfi_high_2013}. 
Monolayer MoSe$_2$ flakes are exfoliated from bulk 2H-MoSe$_2$ crystals on Nitto Denko tape \cite{novoselov2005two} and then exfoliated again on a polydimethylsiloxane (PDMS) stamp placed on a glass slide for inspection under the optical microscope, see supplement for further details.

\section{Supplement}
\subsection{Nano-structure fabrication and TMD exfoliation} \label{partFab}

The silicon pillars were fabricated in a top-down approach via electron-beam lithography (EBL) and subsequent anisotropic plasma etching \cite{han_realization_2011, guerfi_high_2013}.
A negative-tone resist (hydrogen silsesquioxane, HSQ) was spin-coated as thin film (around 50nm) on a commercial silicon-on-insulator (SOI) substrate. 
The SOI consists of bulk silicon followed by a 145nm thick burried oxide layer made of SiO$_2$ (BOX). On top of the BOX follows a 95nm high silicon overlayer, which serves as building material for the nano-structures. 
The nano-patterns were written in the HSQ resist using a RAITH 150 writer at an energy of 30 keV.
After EBL exposure, the resist was developed in \(25\,\)\% tetramethylammonium hydroxide (TMAH). 
Finally, the patterns were etched into the 95nm thick silicon overlayer via reactive ion etching (RIE) in a SF$_6$/C$_4$F$_8$ plasma. RIE was stopped once etching arrived at the BOX layer.
After the processing, an additional SiO$_2$ layer of approximately 30nm height remains on top of the structured Si, which is the residual developped HSQ resist.

Monolayer MoSe$_2$ flakes are exfoliated from bulk 2H-MoSe$_2$ crystals on Nitto Denko tape \cite{novoselov2005two} and then exfoliated again on a polydimethylsiloxane (PDMS) stamp placed on a glass slide for inspection under the optical microscope. Prior to transfer of MoSe$_2$ monolayers, the substrate with the nanoresonators is wet-cleaned by 60 s ultrasonication in acetone and isopropanol and exposed to oxygen-assisted plasma.


\begin{figure*}[h]
\includegraphics[width=1\linewidth]{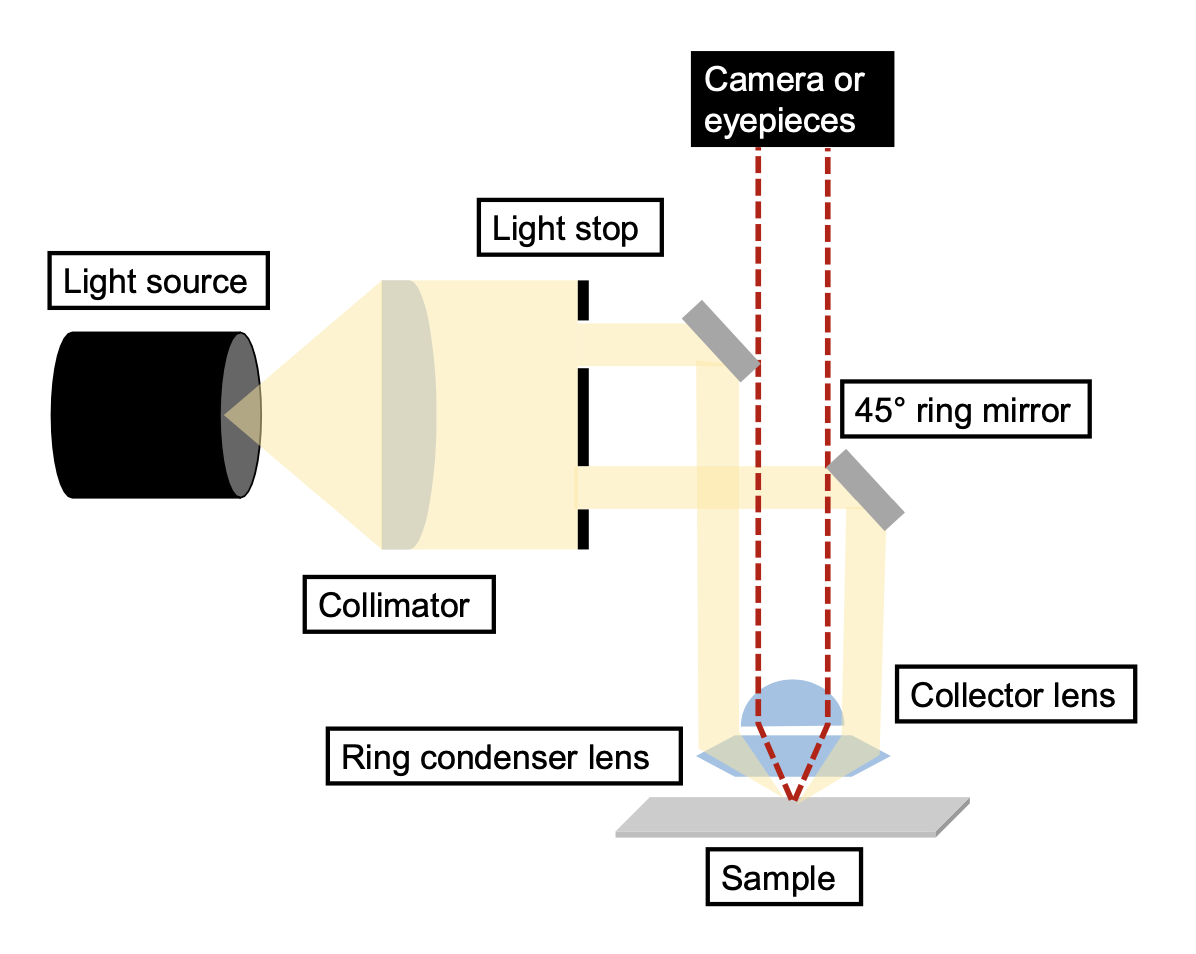}
\caption{\label{fig:figS1} Darkfield setup.}
\end{figure*}


\begin{figure*}[h]
\includegraphics[width=1\linewidth]{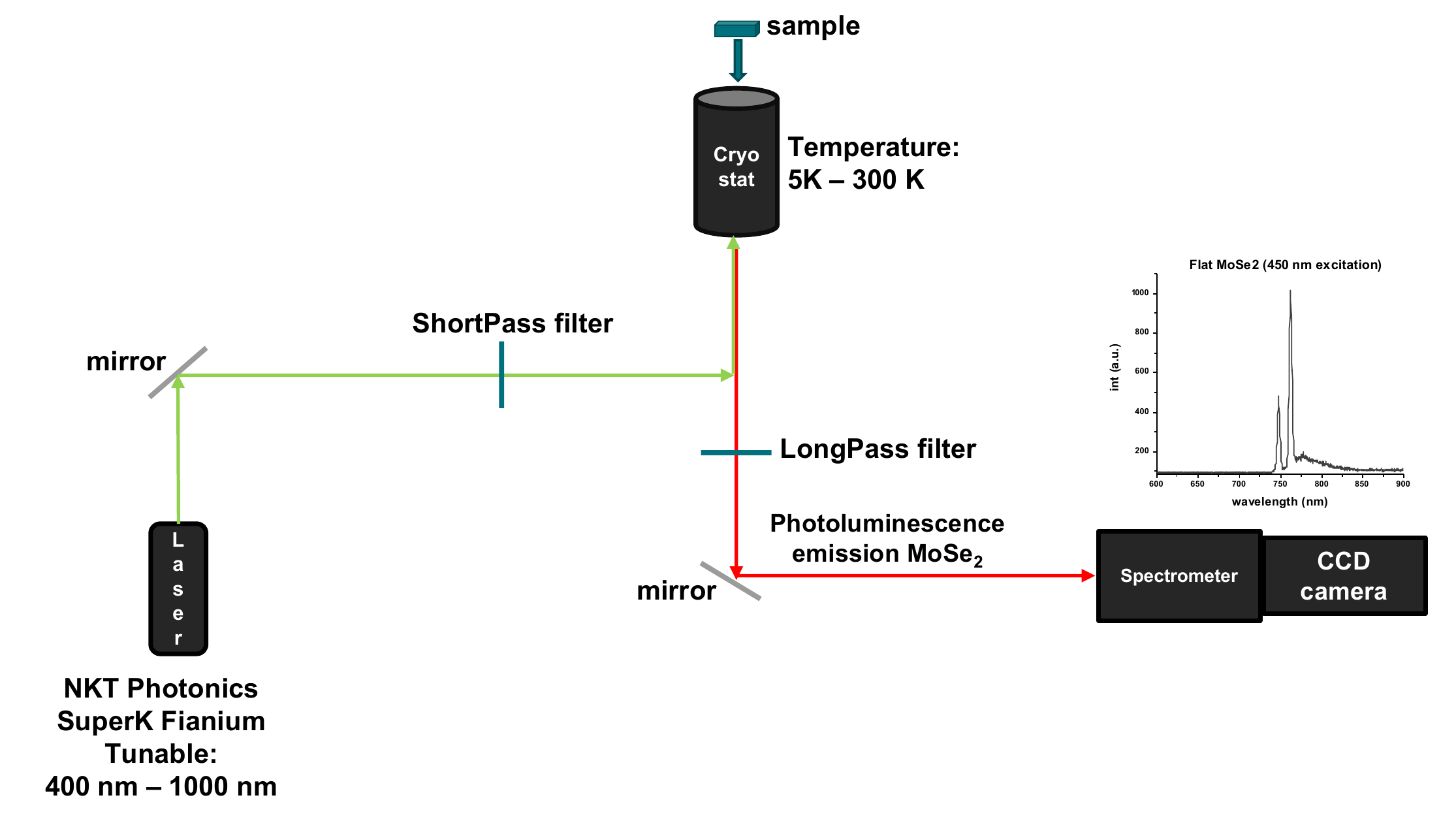}
\caption{\label{fig:figS2} Photoluminescence excitation experiment setup.}
\end{figure*}


\begin{figure*}[h]
\includegraphics[width=1\linewidth]{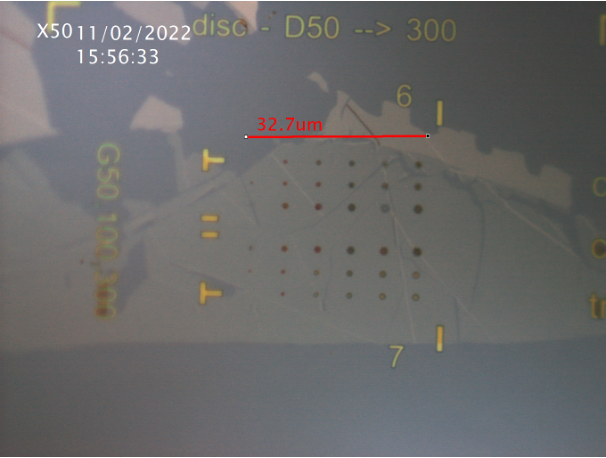}
\caption{\label{fig:figS3} \textbf{Optical bright field image.} Nanoresonators covered with MoSe$_2$ with objective x50.}
\end{figure*}

\clearpage
\subsection{Power Dependence of PL intensity}
\indent To evaluate the enhancement we also carried out power dependent experiments. We keep temperature, position, and excitation wavelength fixed, and now we vary the optical power from 10 - 500~nW (see SI Fig. S4). The enhancement we measure in emission (on the NRs as compared to next to them) is a lower bound for the enhancement in absorption as the generated excitons do not all recombine radiatively (i.e. by emitting a photon in our detection window) \cite{shin2014photoluminescence, kumar2014exciton, zhang2021defect, han2018exciton}.\\
\begin{figure*}[h]
\includegraphics[width=0.6\linewidth]{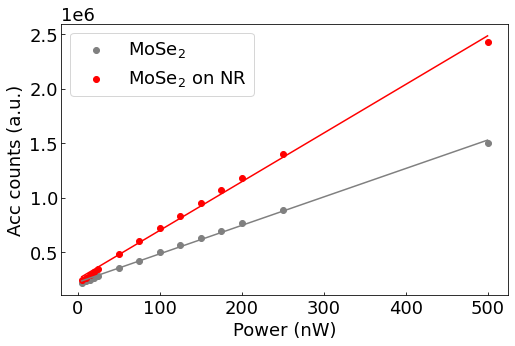}
\caption{\label{fig:figS4} Study of power dependence of MoSe$_2$ on Heptamer D300 G50 (red) and without NR (gray). Fixed excitation energy at 2.25 eV (550~nm), varied the laser power from 5 to 500~nW, and calculated the linear regression. For the red line the slope is 0.509 and for the gray it is 0.414.}
\end{figure*}


\begin{figure*}[h]
\includegraphics[width=1\linewidth]{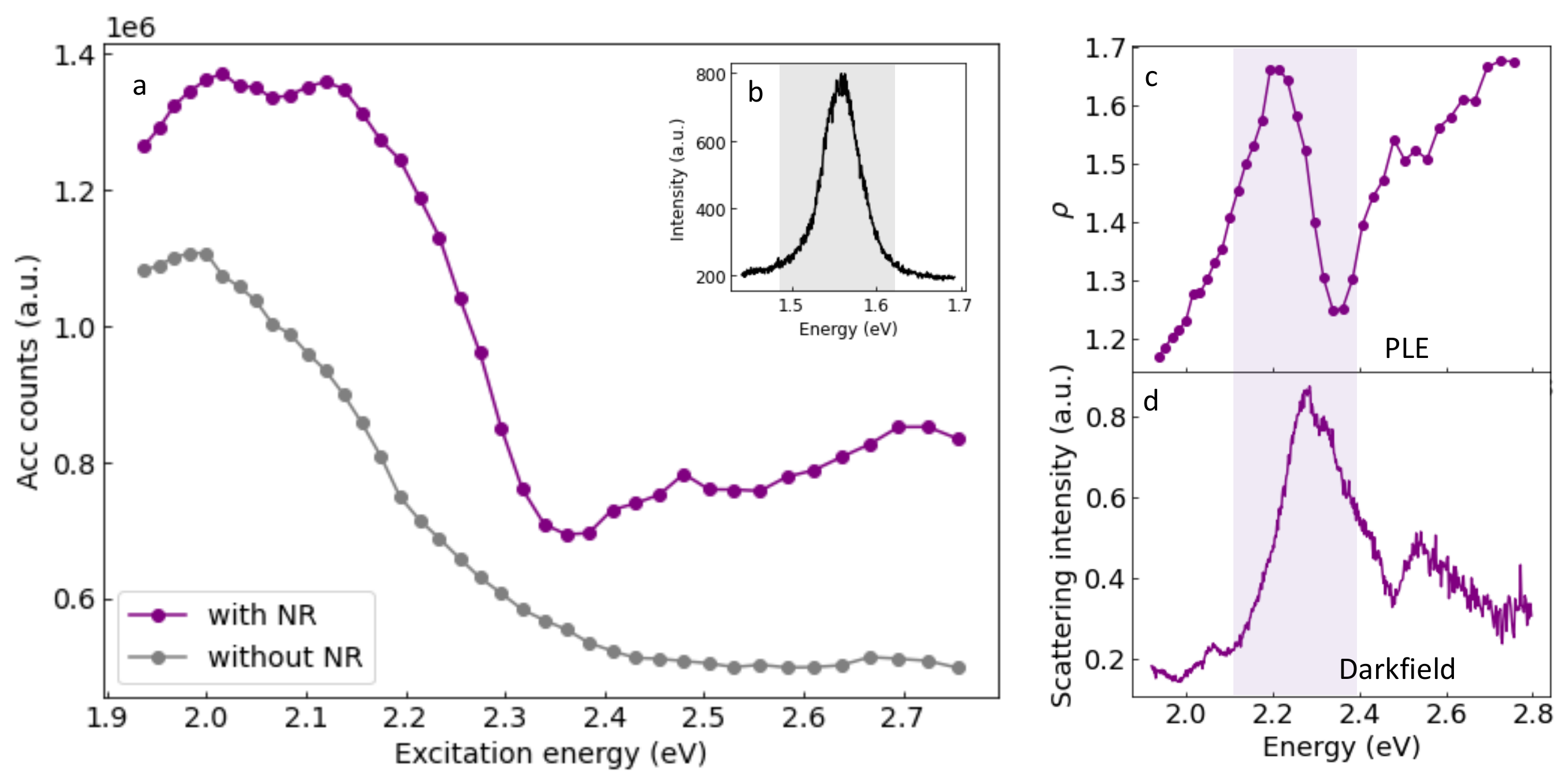}
\caption{\label{fig:figS5} Room temperature results of MoSe$_2$ on hexamer D300 G100. a) Photoluminescence excitation. b) PL of MoSe$_2$ at 295~K, c) $\rho$ ratio as defined in the main text of the Pl intensity with and without the nanoresonator interaction}
\end{figure*}


\begin{figure*}[h]
\includegraphics[width=1\linewidth]{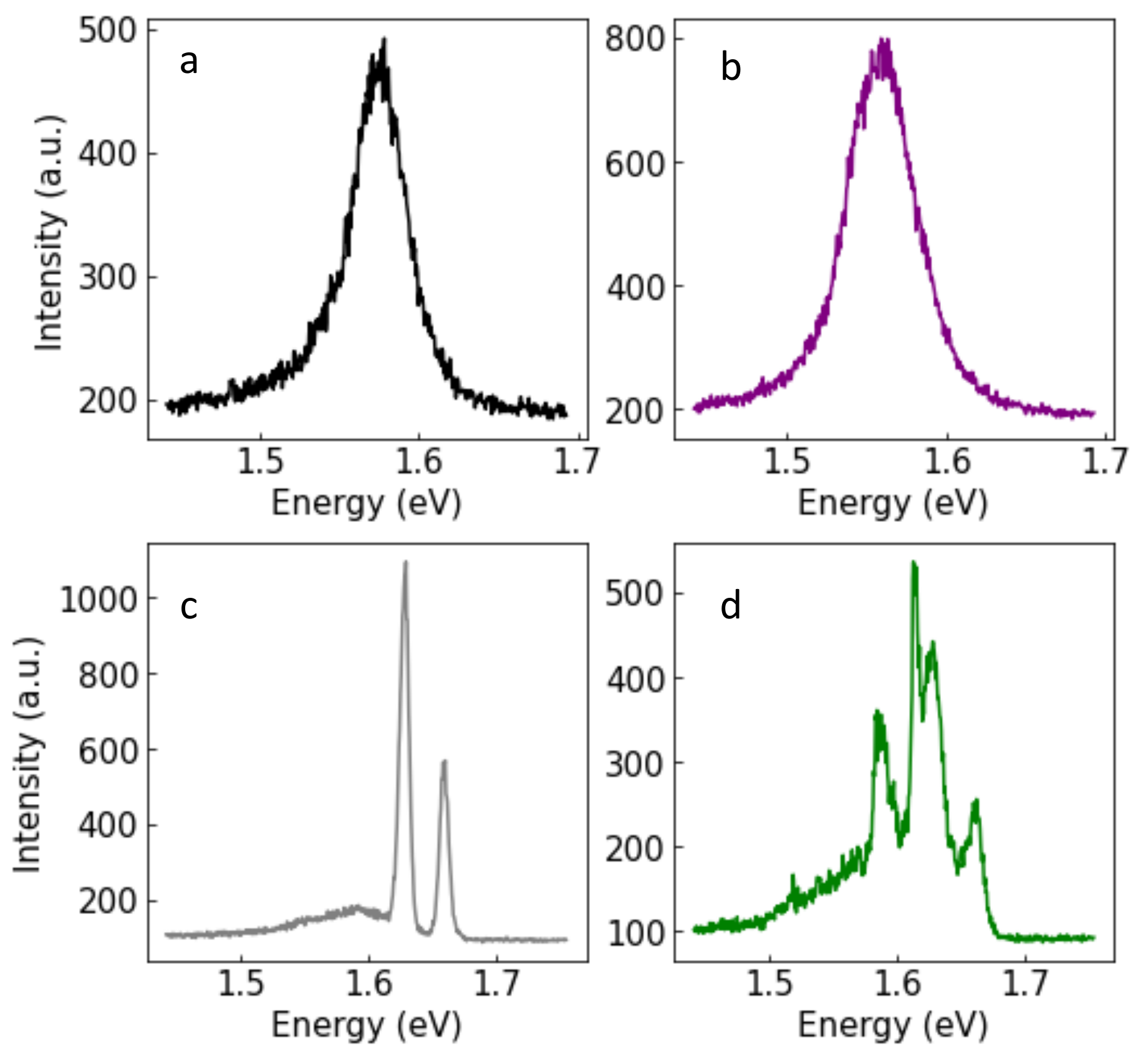}
\caption{\label{fig:figS6} Examples of PL from experiments carried out at 520 nm excitation (2.47 eV) a) MoSe$_2$ at 295 K, b) MoSe$_2$ at 295 K on top of NR Hexamer D300 G100, c) MoSe$_2$ at 5 K, d) MoSe$_2$ at 5 K on top of NR Heptamer D300 G50.}
\end{figure*}

\clearpage
\subsection{Simulations details}

\begin{figure*}[h]
\includegraphics[width=0.9\linewidth]{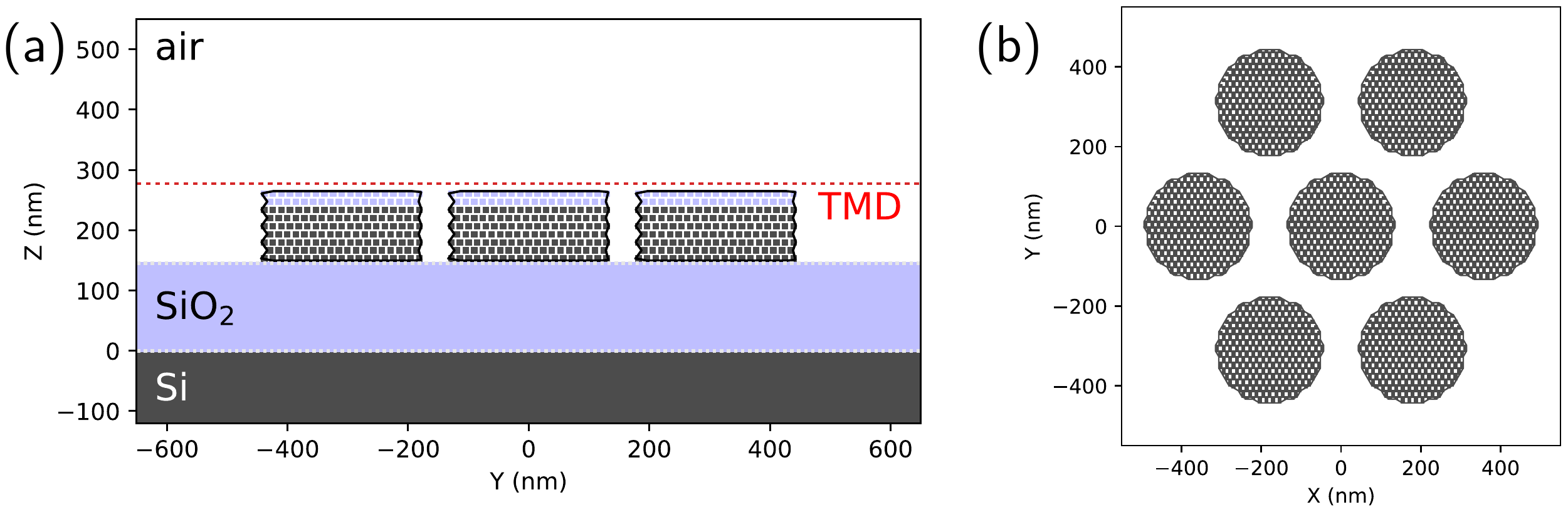}
\caption{\label{fig:figS7} Geometry used for the GDM simulations. (a) side view, (b) top view. An 145nm SiO$_2$ layer lies on a bulk silicon substrate. On top of this, the heptamers are discretized with a hexagonal compact mesh using a nominal stepsize of 16nm (corresponds to 13.47nm in $Z$-direction). Each pillar has the same size with a diameter of 250nm or 300nm, a height of 94nm (silicon part) and an SiO$_2$ capping of 28nm thickness. The near-field evaluation plane is at half of a stepsize above the pillars, as indicated in (a) by a red line (``TMD''). }
\end{figure*}

\begin{figure*}[h]
\includegraphics[width=0.9\linewidth]{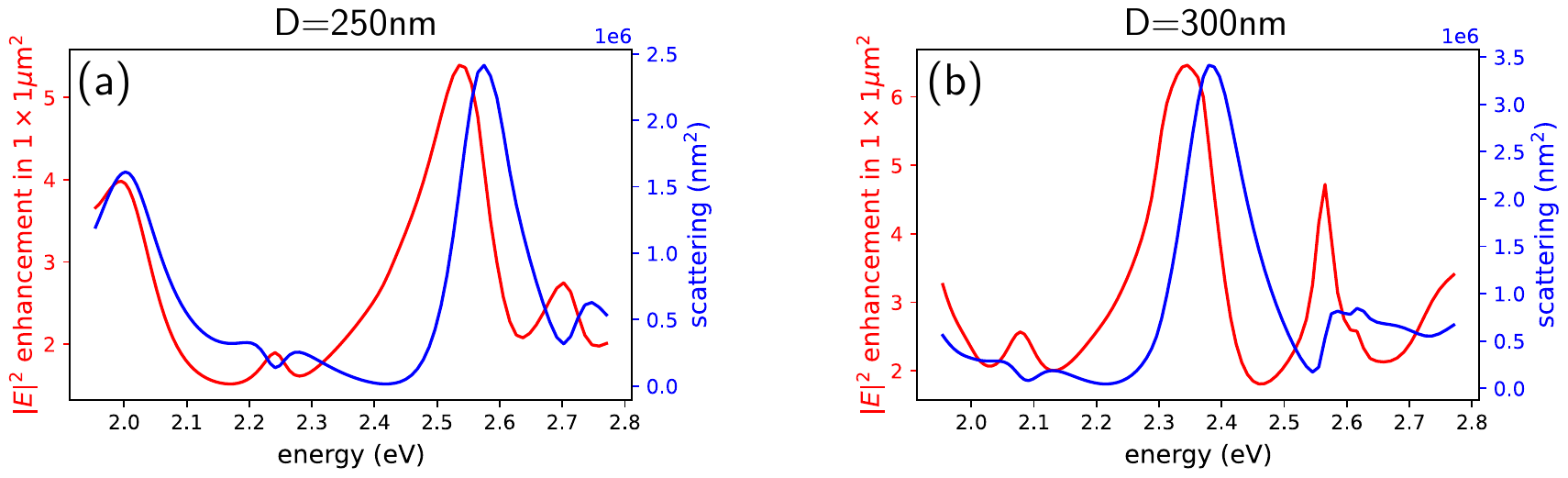}
\caption{\label{fig:figS8} Average near-field enhancement (red) vs. far-field scattering spectra (blue) for (a) the $D=250$nm and (b) the $D=300$nm heptamer. The near-field resonance shift is consistent with the experimental observation.}
\end{figure*}


\begin{figure*}[h]
\includegraphics[width=0.9\linewidth]{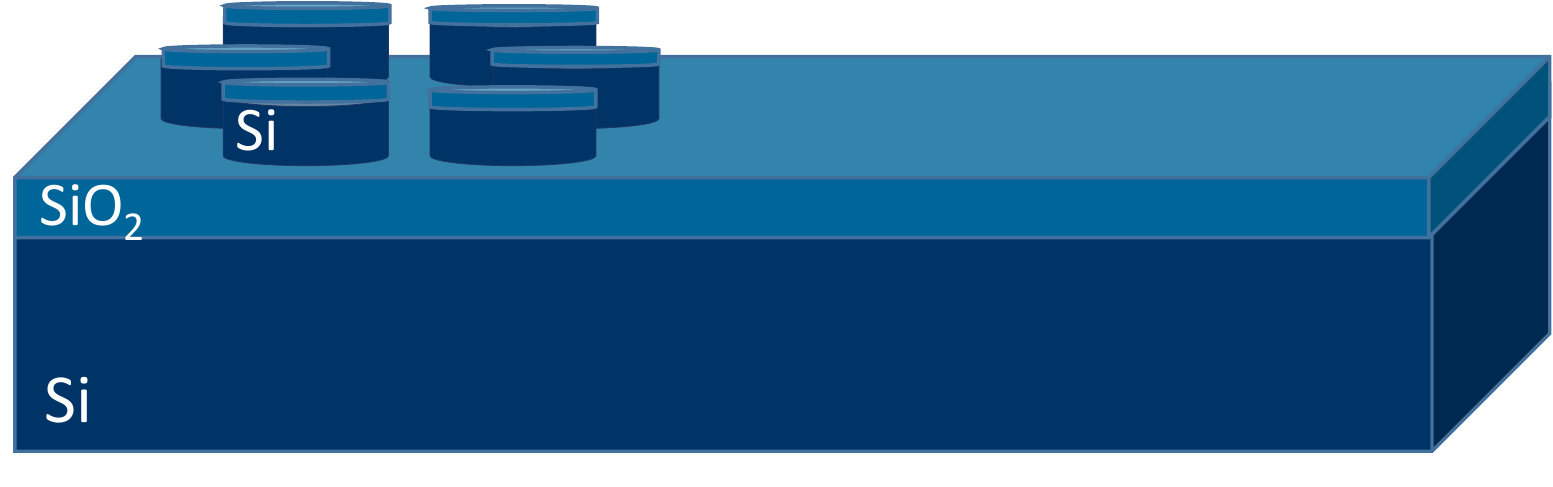}
\caption{\label{fig:figS9} Side view of the substrate with MoSe$_2$ on top. From top to bottom the first SiO$_2$ layer is 30~nm of height, Si pillars are 95~nm + SiO$_2$ 30~nm, SiO$_2$ layer is 145~nm thick, everything on a Si substrate.}
\end{figure*}


\begin{figure*}[h]
\includegraphics[width=0.9\linewidth]{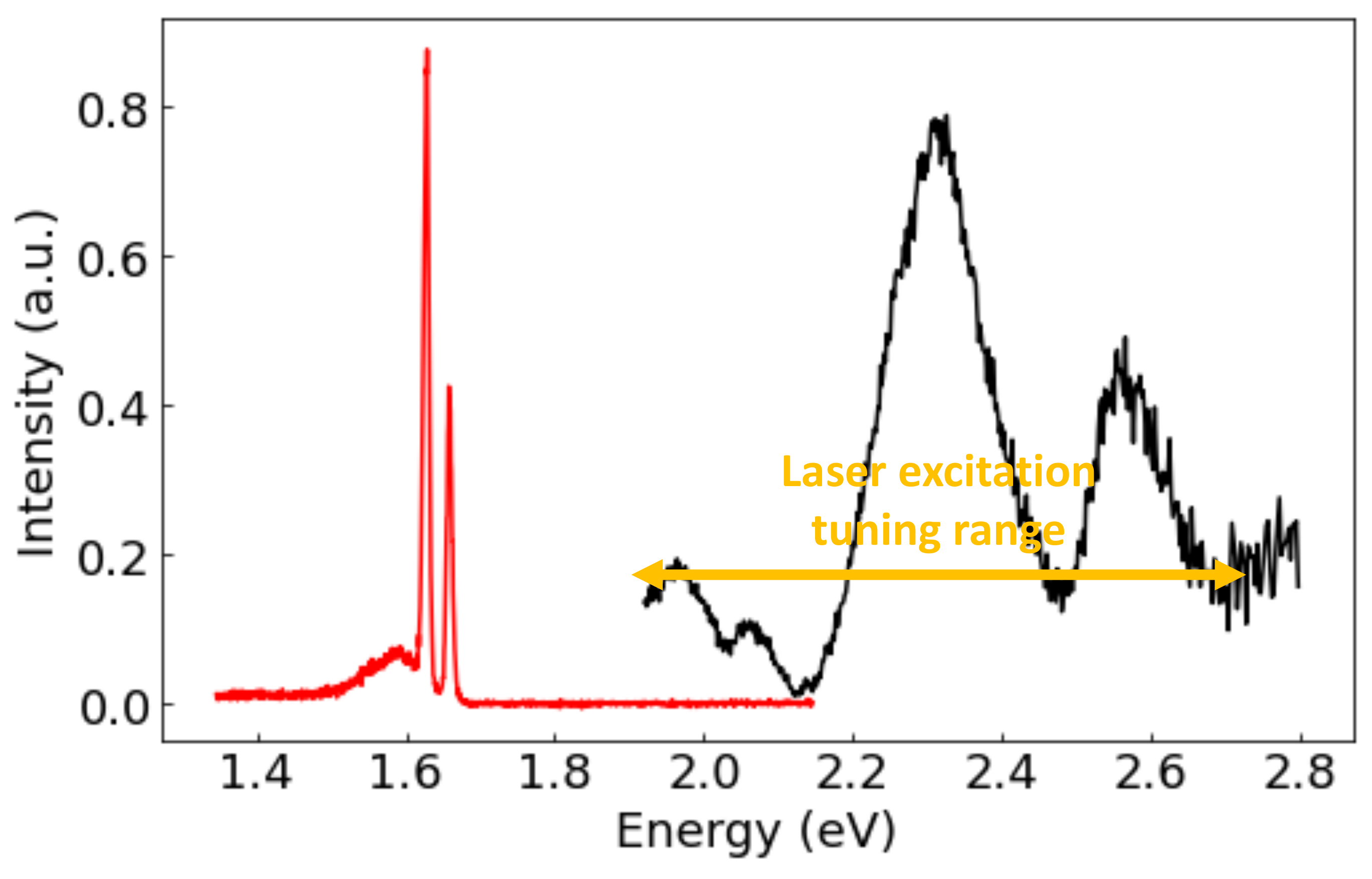}
\caption{\label{fig:figS8} In red MoSe$_2$ emission at 5~K with excitation of 520~nm, in black the darkfield scattering intensity of a nanoresonator (NR) with 7 pilars (heptamer) diameter 300~nm and gap between pilars 100~nm. The yellow arrow represents the range of energies used to excite the sample [450~nm - 650~nm] with our tunable supercontinuum laser.}
\end{figure*}


\begin{figure*}[h]
\includegraphics[width=0.6\linewidth]{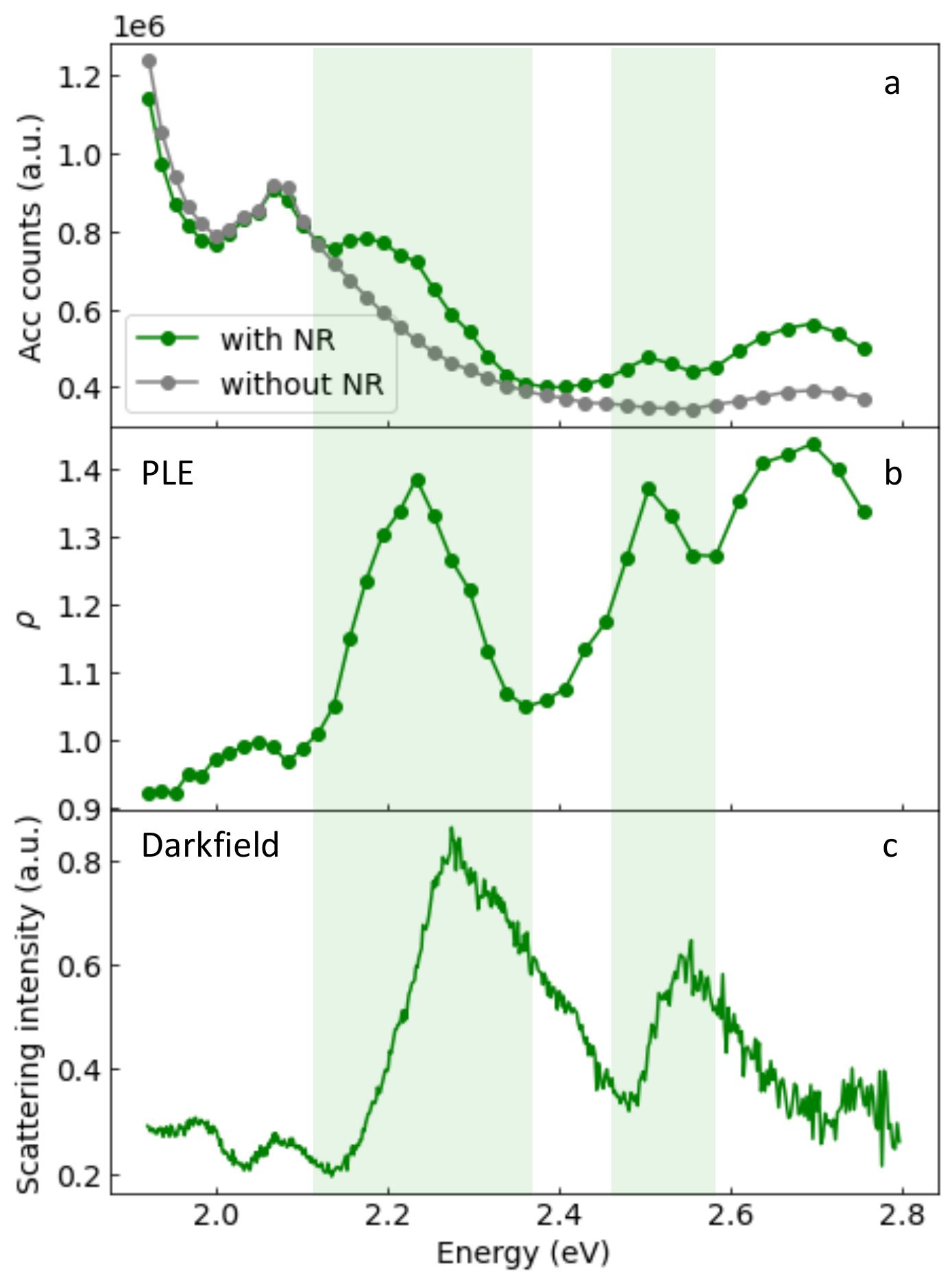}
\caption{\label{fig:figS9} \textbf{Mie resonances with photoluminescence excitation.} Results of measurements taken over nanoresonator with diameter D300 and gap G50 (NR3) (a) Accumulated counts of the photoluminescence (PL) spectra measured on bare MoSe$_2$ (gray) and MoSe$_2$ on top of the nanoresonator (green) excited with a continuum laser at 40 different energies and at T=5~K. (b) The result of dividing PLE over the nanoresonator by PLE on bare MoSe$_2$ $\rho$. (c) the measured darkfield scattering intensity of NR3}
\end{figure*}

\clearpage
\textbf{Data availability statement : } The data that support the findings of this study are available from the corresponding authors upon request.

\textbf{Acknowledgements :} We acknowledge partial funding from ANR HiLight, NanoX project 2DLight, the Institute of quantum technology in Occitanie IQO and a UPS excellence PhD grant. This work was supported by the Toulouse HPC CALMIP (grant p20010), and by the LAAS-CNRS micro and nanotechnologies platform, a member of the French RENATECH network. IP acknowledges financial support by the Hellenic Foundation for Research and Innovation (H.F.R.I.) under the “3rd Call for H.F.R.I. Research Projects to support Post-Doctoral Researchers” (Project Number: 7898).

\textbf{Author contributions :} V.L. fabricated the high quality Si-SiO$_2$ starting structure. J.M. and G.L. fabricated the nanoresonators based on the design of A.C.,G.A. and V.P.  A.E-R. and I.P. deposited the TMD layer on the nanoresonators. A.E-R., P.R.W. and J-M.P. performed dark-field experiments that were analyzed with all co-authors. D.L. and X.M. mounted the laser system for PLE. A.E-R. performed PLE and analyzed data with I.P.  P.R.W., A.E-R., A.G. and A.C. performed modelling of the resonances. A.E-R. wrote the manuscript with input from all co-authors. V.P., G.L., P.R.W. and B.U. supervised the project.

\textbf{Competing interests :} The authors declare no competing interests.

\clearpage

\end{document}